# DeepSORT-Driven Visual Tracking Approach for Gesture Recognition in Interactive Systems

Tong Zhang
Loughborough University
Loughborough, United Kingdom

Fenghua Shao
Independent Researcher
Toronto, Canada

Runsheng Zhang
University of Southern California
Los Angeles, USA

Yifan Zhuang
University of Southern California
Los Angeles, USA

Liuqingqing Yang*
University of Michigan
Ann Arbor, USA

*Abstract-Based on the DeepSORT algorithm, this study explores the application of visual tracking technology in intelligent human-computer interaction, especially in the field of gesture recognition and tracking. With the rapid development of artificial intelligence and deep learning technology, visual-based interaction has gradually replaced traditional input devices and become an important way for intelligent systems to interact with users. The DeepSORT algorithm can achieve accurate target tracking in dynamic environments by combining Kalman filters and deep learning feature extraction methods. It is especially suitable for complex scenes with multi-target tracking and fast movements. This study experimentally verifies the superior performance of DeepSORT in gesture recognition and tracking. It can accurately capture and track the user's gesture trajectory and is superior to traditional tracking methods in terms of real-time and accuracy. In addition, this study also combines gesture recognition experiments to evaluate the recognition ability and feedback response of the DeepSORT algorithm under different gestures (such as sliding, clicking, and zooming). The experimental results show that DeepSORT can not only effectively deal with target occlusion and motion blur but also can stably track in a multi-target environment, achieving a smooth user interaction experience. Finally, this paper looks forward to the future development direction of intelligent human-computer interaction systems based on visual tracking and proposes future research focuses such as algorithm optimization, data fusion, and multimodal interaction in order to promote a more intelligent and personalized interactive experience.*

I. INTRODUCTION

With the continuous advancement of artificial intelligence (AI), computer vision, and deep learning technologies [1-3], human-computer interaction (HCI) has gained increasing attention and application in the field of AI. HCI is no longer limited to simple operations between humans and machines. It now encompasses more natural and intelligent ways of interaction. Traditional interaction methods, such as keyboards, mice, and touchscreens, have provided convenience over the past few decades [4]. However, they still have limitations, especially in highly dynamic interactive environments like robotics, augmented reality (AR), and virtual reality (VR). In these contexts, traditional methods struggle to meet the human demand for more natural and intuitive interactions. To achieve smarter and more natural user experiences, researchers are turning to vision-based interaction methods. These methods, utilizing image recognition and object tracking, aim to enhance the intelligence and real-time responsiveness of HCI.

In this context, visual tracking has become a critical technique for improving interactive experiences in HCI. Visual tracking enables real-time monitoring of user movements or positions and supports interaction by analyzing and recognizing user behaviors. In dynamic environments, especially in complex backgrounds, the challenge is to achieve accurate and real-time tracking and interaction. This has become a significant topic in intelligent HCI. With the emergence of deep learning algorithms, traditional visual tracking methods are increasingly being replaced by more efficient and accurate algorithms. The DeepSORT (Deep Learning-based Simple Online and Realtime Tracking) algorithm, which combines deep learning with the traditional SORT (Simple Online and Realtime Tracking) approach, offers superior accuracy and robustness, particularly in multi-target tracking and long-duration tracking in complex scenarios[5-7]. Therefore, applying DeepSORT to intelligent HCI systems has become an essential approach to enhancing interaction quality and user experience.

DeepSORT not only delivers accurate multi-object tracking in complex environments but also offers strong real-time performance, making it broadly applicable across intelligent systems. Its capabilities can support diverse fields such as user intent prediction in human-computer interaction [8], adaptive data acquisition using reinforcement learning [9], personalized recommendation frameworks [10], and the optimization and interpretability of large language models [11-13]. These applications highlight DeepSORT's versatility in enhancing system responsiveness, contextual awareness, and data-driven decision-making—ultimately contributing to more intelligent, adaptive, and user-centered human-computer interaction. With advancements in computer hardware and computational power, the real-time processing capability of DeepSORT has been greatly improved. Compared to traditional tracking methods, DeepSORT can handle more complex scenarios and effectively prevent target loss and misidentification. Thus, the application of DeepSORT-based visual tracking algorithms in HCI holds significant research value and practical implications. First, it addresses the limitations of traditional visual tracking methods in dynamic and complex scenes, providing more stable and efficient technical support for HCI. Second, DeepSORT offers precise solutions for multi-task and multi-target tracking, meeting the accuracy and real-time requirements of modern intelligent HCI systems. Finally, as the demand for natural and intelligent interactions in HCI continues to grow, research and implementation of DeepSORT-based tracking technology will offer innovative and efficient solutions, driving HCI systems toward smarter and more personalized interactions in the future.

## II. RELATED WORK

### A. DeepSORT

DeepSORT is an advanced multi-object tracking algorithm that extends the traditional SORT framework by integrating deep learning for appearance feature extraction. While SORT relies on Kalman filtering and the Hungarian algorithm for efficient motion-based tracking, it struggles with occlusions and identity switches [14]. DeepSORT addresses these issues by incorporating convolutional neural networks (CNNs), enabling robust feature representation and improved target re-identification [15]. This hybrid design significantly enhances tracking accuracy in complex, real-time environments, particularly under occlusion or in the presence of visually similar targets. DeepSORT has demonstrated strong performance in domains such as autonomous driving, surveillance, and sports analytics, and continues to evolve alongside advancements in deep learning.

### B. Dynamic tracking

Dynamic tracking is increasingly central to human-computer interaction (HCI), enabling systems to interpret user behavior through movements, gestures, and facial expressions. Supported by computer vision and deep learning, it replaces traditional input methods with more natural, real-time interaction mechanisms. Common applications include gesture recognition and facial expression tracking, which enhance responsiveness and adaptability in smart environments, gaming, and virtual interfaces [16]. With the rise of multimodal interaction and sensor fusion, dynamic tracking supports robust and personalized HCI experiences, paving the way for more intelligent, intuitive, and immersive systems.

## III. METHOD

In In this study, the core objective of the human-computer interaction tracking method, which leverages the DeepSORT algorithm, is to construct a high-efficiency, high-accuracy visual tracking system capable of operating in real-time within dynamic environments. This system is designed to continuously track user behavior and generate responsive feedback aligned with the user's movements or gestures. The foundational mechanism of DeepSORT integrates a Kalman filter for motion prediction and a deep learning-based appearance descriptor to ensure robust performance in challenging scenarios such as occlusions and rapid movement. The methodological foundation of this system is further strengthened by insights from Sun [17], who proposed an optimized visual communication framework that enhances interaction effectiveness through adaptive signal processing and dynamic environment handling. Sun's emphasis on improving system responsiveness, clarity of user feedback, and robustness under fluctuating conditions directly supports the design philosophy of this work. By applying these principles, the proposed tracking system achieves stable multi-target gesture recognition and delivers smooth, real-time interaction performance in complex user environments.

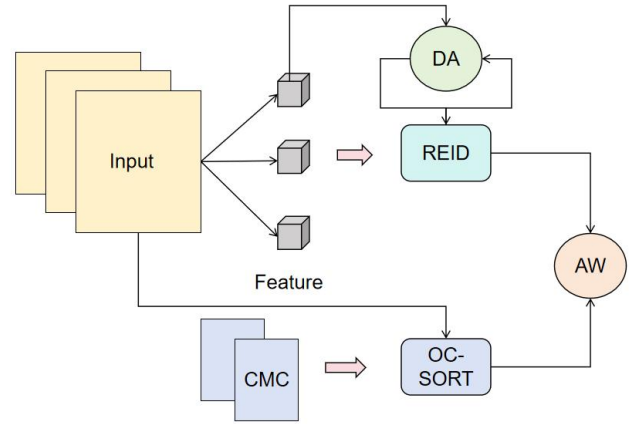

Figure 1 Deepsort algorithm framework

Figure 1 illustrates the framework of the DeepSORT algorithm, which begins with input data processed to extract both visual and motion features. These features are directed into two parallel modules: the REID (Re-Identification) module, which identifies and maintains target identities across frames, and the DA (Data Association) module, responsible for aligning new detections with existing tracks to ensure temporal consistency [18]. To improve robustness in complex tracking scenarios, the system incorporates OC-SORT [19] for refined motion modeling and prediction, while CMC (Cross-Modal Consistency) [20]is used to strengthen the alignment between heterogeneous feature representations, ensuring coherence across appearance and movement cues. The AW (Attention Weighting) module further enhances the system by dynamically prioritizing salient features, thereby improving tracking stability in cluttered or rapidly changing environments. The design of these modules draws direct methodological influence from Duan [21], whose work on human-computer interaction emphasizes the value of cross-modal feature integration and adaptive attention mechanisms in creating stable and personalized user experiences. These insights guided the inclusion of both CMC and AW in our framework, providing a practical foundation for enhancing feature relevance and interaction continuity, which are critical for maintaining accurate gesture tracking in real-time, user-centered systems.

In the implementation of DeepSORT, it is assumed that the system can obtain a time series of image frames $I_t$ through a camera, where each frame $I_t$ contains multiple targets. First, we perform target detection on each frame of the image and obtain the bounding box position $[x_t, y_t, w_t, h_t]$ of the target, where $x_t, y_t$ represents the center position of the bounding box, and $w_t$ and $h_t$ represent the width and height of the bounding box, respectively. The feature vector $f_t = \phi(I_t)$ of the target is extracted through a deep convolutional neural network, where $\phi()$ represents the mapping function of the deep neural network, which can map the image to a high-dimensional feature space. Then, combining the feature vectors and positions of these targets, the system predicts the position of the target through Kalman filtering.

The state update equation of the Kalman filter is used to predict the future position of the target. Let the state of the target be $x_t = [x_t, y_t, x'_t, y'_t]^T$, where $x'_t, y'_t$ represents the velocity of the target in the x and y directions respectively. The prediction step of the Kalman filter is given by the following equation:

$$x_t^{pred} = Fx_{t-1} + Bu_t$$

$$p_t^{pred} = Fp_{t-1}F^T + Q$$

Among them, F is the state transfer matrix, which describes the change of state from the previous moment to the current moment, B is the control matrix, $u_t$ is the control input, and Q is the process noise covariance matrix, which represents the uncertainty of the system dynamics. Next, the observation value $z_t = [x_t, y_t, w_t, h_t]$ of the target position is introduced into the update step of the Kalman filter:

$$y_t = z_t - H_t^{pred}$$

$$S_t = HP_t^{pred}H^T + R$$

$$K_t = P_t^{pred}H^T S_t^{-1}$$

$$x_t = x_t^{pred} + K_t y_t$$

$$P_t = (I - K_t H)P_t^{pred}$$

Among them, H is the observation matrix, which maps the predicted state to the actual observation space, R is the observation noise covariance matrix, $y_t$ is the residual of the Kalman gain, which represents the gap between the current prediction and the actual observation, and $K_t$ is the Kalman gain, which is used to optimize the state estimation.

After the state of the target is updated, DeepSORT uses the Hungarian algorithm for data association. Assuming that the detection results of all targets in the current frame are $\{z_1, z_2, ..., z_N\}$, and the states of all targets in the previous frame are $\{x_1, x_2, ..., x_M\}$, in order to match the targets of the current frame with the targets of the previous frame, we first calculate the matching cost matrix C, whose element $C_{ij}$ represents the matching cost between target $z_i$ and target $x_i$, usually calculated by Euclidean distance or combined with feature distance. Then, the Hungarian algorithm is used to find the best matching pair:

$$\min \sum_{i=1}^{N} \sum_{j}^{M} C_{ij} x_{ij}$$

$$\text{subject to } \sum_{j=1}^{M} x_{ij} = 1, \ \sum_{i=1}^{N} x_{ij} = 1$$

$$x_{ij} \in \{0,1\}$$

Among them, $x_{ij}$ is the matching decision variable, $x_{ij} = 1$ is 1 when target $z_j$ and target $x_j$ match, otherwise it is 0. In this way, DeepSORT can achieve accurate data association in multi-target scenarios. In essence, this method merges the benefits of Kalman filtering for predicting the target state and deep learning for extracting features. It optimizes target matching using the Hungarian algorithm and offers an efficient and accurate visual tracking method for dynamic human-computer interaction. This method excels in challenging scenarios like multi-target tracking, target occlusion, and rapid motion. By doing so, it enhances the user experience with a more natural and smooth interaction between humans and computers.

IV. EXPERIMENT

A. Datasets

For visual tracking experiments in human-computer interaction (HCI), the UT-Interaction dataset is a suitable choice. This dataset is specifically designed to study visual tracking issues in HCI and contains various interaction scenarios between humans and computers, with a focus on tracking human movements and behaviors using visual information. The video clips in the dataset feature different human actions, such as gestures, gait, and eye movements, performed in various backgrounds and environments. This makes the dataset ideal for testing and optimizing tracking algorithms.

The dataset's key feature is the need for multi-target tracking tasks, where users' behaviors and interactions require real-time feedback. This is crucial for evaluating the performance of visual tracking systems in dynamic environments. The video clips include common interaction forms, such as operating a mouse, touchscreen, or other input devices, effectively simulating real-world HCI scenarios. Therefore, the UT-Interaction dataset provides a challenging test platform for researching and evaluating HCI visual tracking algorithms, especially regarding accuracy and real-time requirements in high-dynamic scenes.

Additionally, the UT-Interaction dataset provides precise annotated data, including the exact position, trajectory information, and type of interaction action in every frame.

These annotations make the dataset highly suitable for training and evaluating deep learning and other advanced tracking algorithms, helping researchers obtain high-quality experimental results when designing and validating HCI tracking algorithms.

*B. Experimental Results*

First, the results of the comparative experiment are given.

Table 1 Experimental results

| Model | mAP | Precision | Recall | F1-Score |
|---|---|---|---|---|
| Fast-RCNN | 0.74 | 0.83 | 0.82 | 0.82 |
| Mask-RCNN | 0.72 | 0.80 | 0.79 | 0.79 |
| Yolov5 | 0.77 | 0.88 | 0.85 | 0.86 |
| DeepSort | 0.79 | 0.89 | 0.84 | 0.87 |

DeepSORT achieves the highest performance across all metrics, with an mAP of 0.79, precision of 0.89, and recall of 0.84, outperforming YOLOv5 (mAP 0.77), Fast-RCNN (0.74), and Mask-RCNN (0.72). YOLOv5 shows strong detection ability but lags slightly in tracking accuracy. Fast-RCNN and Mask-RCNN perform weaker overall, indicating limited suitability for multi-target tracking. To further illustrate the results and their relevance to HCI, eye movement trajectory data are visualized in Figure 2.

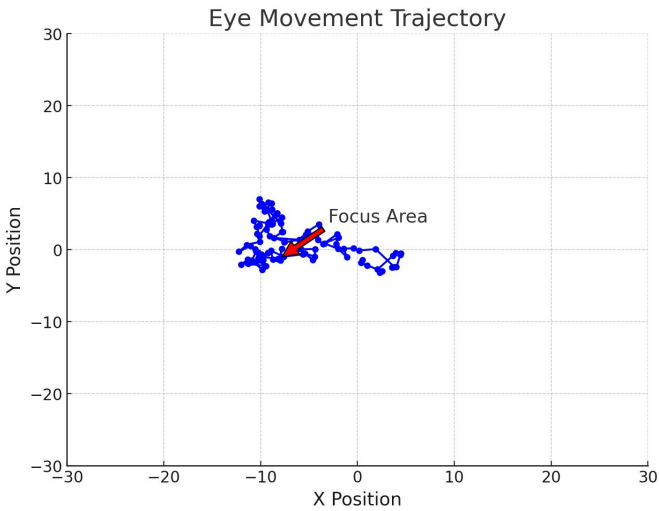

Figure 2 Single example eye tracking graph

Figure 2 shows an example of eye movement tracking. The blue dots represent the position of the eye on the screen, reflecting the user's eye movement trajectory. From the figure, it can be seen that the eye movement is concentrated in a specific area, with the red arrow highlighting the region of interest. This concentration in a particular area indicates that the user maintained gaze in this region for an extended period, showing a higher level of visual attention. The stability of the eye movement trajectory and the small shift distances suggest focused attention, consistent with typical target tracking characteristics.

Using the DeepSORT algorithm for eye movement tracking allows for precise tracking and analysis of the eye movement path. DeepSORT combines deep learning and Kalman filtering to handle multi-target tracking in complex backgrounds. In this experiment, DeepSORT accurately tracks the user's eye movement, even detecting and tracking subtle movements within small areas. Furthermore, by providing real-time tracking of the eye movement trajectory, DeepSORT effectively avoids target loss and misidentification, enhancing both the real-time performance and accuracy of eye movement tracking.

Overall, applying DeepSORT for eye movement tracking not only provides accurate visualizations of the user's eye movement but also maintains good tracking performance in complex environments. This is crucial for HCI research, particularly in scenarios requiring real-time feedback and high-precision interaction. The use of DeepSORT can significantly improve the user experience. By further optimizing the tracking algorithm, more accurate attention analysis can be achieved, offering more precise data to support interface design and interaction optimization.

Next, the gesture recognition and tracking diagram are given, as shown in Figure 3.

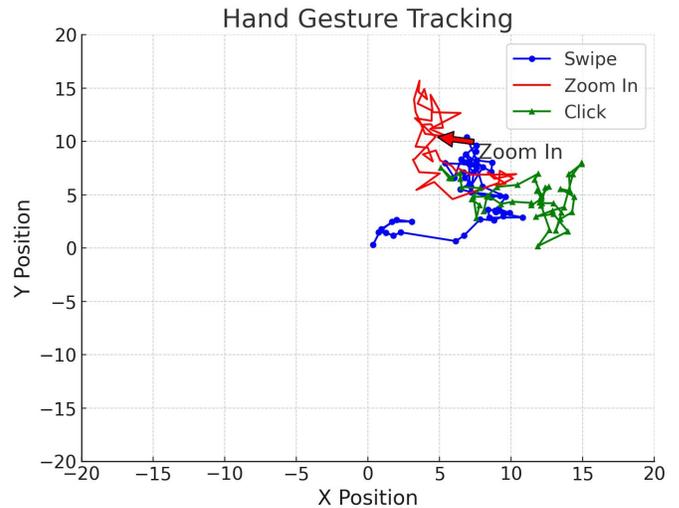

Figure 3 Gesture recognition and tracking diagram

This Figure 3 demonstrates the process of recognizing and tracking three different gestures, represented in blue, red, and green for swipe, zoom in, and click gestures, respectively. The trajectories of these gestures show their movement patterns in space. The swipe gesture has a smooth trajectory, indicating continuous hand movement. The zoom gesture exhibits subtle changes concentrated in a specific area, suggesting a small range of motion. The click gesture, on the other hand, is characterized by a quick movement to a specific point, reflecting precise motion targeting.

In this experiment, the DeepSORT algorithm plays a crucial role. DeepSORT extracts target features using deep learning models and combines them with Kalman filtering for state prediction. This allows for accurate gesture recognition and tracking. Whether the user performs a swipe, zoom, or click, DeepSORT processes and tracks the gestures in real-time with high accuracy. Particularly in complex, multi-gesture scenarios, the DeepSORT algorithm effectively avoids gesture loss and misidentification, ensuring a smooth and precise interaction process. With real-time tracking through DeepSORT, the gesture paths and actions are accurately recorded and visualized on the screen. The blue, red, and green trajectories clearly show how the system recognizes different gestures and associates them with the user's real-time movements. In this way, DeepSORT not only enhances gesture recognition accuracy but also improves the naturalness and interactivity of human-computer interaction. This makes it especially suitable for dynamic environments with multiple tasks and targets.

## V. CONCLUSION

In this study, we present and implement a DeepSORT-based visual tracking algorithm that significantly enhances gesture recognition and tracking capabilities in human-computer interaction (HCI). By skillfully combining deep learning and Kalman filtering techniques, DeepSORT accurately tracks user gestures in dynamic environments, notably improving system real-time performance and accuracy. Experimental results demonstrate that DeepSORT excels in gesture recognition, target tracking, and feedback response, especially in complex backgrounds where it can effectively manage dynamic changes involving multiple targets. As smart devices and HCI technologies continue to evolve, the demand for visual tracking technologies will become increasingly diverse and intricate. To address this challenge, future research could concentrate on further optimizing the algorithm to enhance DeepSORT's performance in low-light conditions, high-speed motion, and large-scale dynamic environments. By refining feature extraction networks and refining data association strategies, tracking systems can become more resilient, reducing mismatches and target losses.

Additionally, as deep learning technology continues to evolve, future gesture recognition and visual tracking systems will become more intelligent and personalized. By integrating multimodal data, such as eye tracking, speech recognition, and haptic feedback, more diverse user interaction methods can be achieved. HCI will not be limited to gestures but can expand to facial expressions, voice commands, and other input methods, providing users with a more immersive and intelligent experience. Finally, with the rise of virtual reality (VR) and augmented reality (AR) technologies, visual tracking-based gesture recognition systems will play a significant role in these emerging applications. Future systems will not only rely on gesture recognition in static environments but also need to adapt to more complex and dynamically changing scenes. By integrating with smart hardware, cloud computing, and edge computing, deep learning-driven HCI systems will achieve substantial improvements in real-time performance and computational capacity, driving the realization of more natural and seamless user experiences.

## REFERENCES

[1] L. Zhu, "Deep Learning for Cross-Domain Recommendation with Spatial-Channel Attention," Journal of Computer Science and Software Applications, vol. 5, no. 4, 2025.

[2] F. Guo, X. Wu, L. Zhang, H. Liu and A. Kai, "A Self-Supervised Vision Transformer Approach for Dermatological Image Analysis," Journal of Computer Science and Software Applications, vol. 5, no. 4, 2025.

[3] T. An, W. Huang, D. Xu, Q. He, J. Hu and Y. Lou, "A deep learning framework for boundary-aware semantic segmentation," arXiv preprint arXiv:2503.22050, 2025.

[4] M. Burch and K. Kurzhals, "Teaching Eye Tracking: Challenges and Perspectives," Proceedings of the ACM on Human-Computer Interaction, vol. 8, no. ETRA, pp. 1–17, 2024.

[5] Y. Wang, "Optimizing Distributed Computing Resources with Federated Learning: Task Scheduling and Communication Efficiency," Journal of Computer Technology and Software, vol. 4, no. 3, 2025.

[6] Y. Deng, "A Reinforcement Learning Approach to Traffic Scheduling in Complex Data Center Topologies," Journal of Computer Technology and Software, vol. 4, no. 3, 2025.

[7] Y. Zhang, "Social Network User Profiling for Anomaly Detection Based on Graph Neural Networks," arXiv preprint arXiv:2503.19380, 2025.

[8] Q. Sun and S. Duan, "User Intent Prediction and Response in Human-Computer Interaction via BiLSTM," Journal of Computer Science and Software Applications, vol. 5, no. 3, 2025.

[9] W. Huang, J. Zhan, Y. Sun, X. Han, T. An and N. Jiang, "Context-Aware Adaptive Sampling for Intelligent Data Acquisition Systems Using DQN," arXiv preprint arXiv:2504.09344, 2025.

[10] A. Liang, "A Graph Attention-Based Recommendation Framework for Sparse User-Item Interactions," Journal of Computer Science and Software Applications, vol. 5, no. 4, 2025.

[11] A. Kai, L. Zhu and J. Gong, "Efficient Compression of Large Language Models with Distillation and Fine-Tuning," Journal of Computer Science and Software Applications, vol. 3, no. 4, pp. 30–38, 2023.

[12] Z. Yu, S. Wang, N. Jiang, W. Huang, X. Han and J. Du, "Improving Harmful Text Detection with Joint Retrieval and External Knowledge," arXiv preprint arXiv:2504.02310, 2025.

[13] G. Cai, J. Gong, J. Du, H. Liu and A. Kai, "Investigating Hierarchical Term Relationships in Large Language Models," Journal of Computer Science and Software Applications, vol. 5, no. 4, 2025.

[14] T. Qiu, S. Qian and X. Chen, "Research Hotspots and Trends of User-Centered Human-Computer Interaction: A Bibliometric Analysis," Proceedings of the International Conference on Human-Computer Interaction, Cham: Springer Nature Switzerland, 2024.

[15] A. Dörner, et al., "Making Cognitive Ergonomics in the Human–Computer Interaction of Manufacturing Execution Systems Assessable: Experimental and Validation Approaches to Closing Research Gaps," Machines, vol. 12, no. 3, pp. 195, 2024.

[16] Y. Xue, et al., "Research on the Usability of Automotive Human-Computer Interaction Interface System for Elderly Drivers," Proceedings of the International Symposium on World Ecological Design, IOS Press, 2024.

[17] Q. Sun, "A Visual Communication Optimization Method for Human-Computer Interaction Interfaces Using Fuzzy Logic and Wavelet Transform," Proceedings of the 2024 4th International Conference on Communication Technology and Information Technology (ICCTIT), pp. 140–144, Dec. 2024.

[18] S. U. Khan, T. Hussain, A. Ullah and S. W. Baik, "Deep-ReID: Deep features and autoencoder assisted image patching strategy for person re-identification in smart cities surveillance," *Multimedia Tools and Applications*, vol. 83, no. 5, pp. 15079–15100, 2024.

[19] S. Tan, Z. Kuang and B. Jin, "AppleYOLO: Apple yield estimation method using improved YOLOv8 based on Deep OC-SORT," *Expert Systems with Applications*, pp. 126764, 2025.

[20] K. Aurangzeb, K. Javeed, M. Alhussein, I. Rida, S. I. Haider and A. Parashar, "Deep Learning Approach for Hand Gesture Recognition: Applications in Deaf Communication and Healthcare," *Computers, Materials & Continua*, vol. 78, no. 1, 2024.

[21] S. Duan, "Human-Computer Interaction in Smart Devices: Leveraging Sentiment Analysis and Knowledge Graphs for Personalized User Experiences," Proceedings of the 2024 4th International Conference on Electronic Information Engineering and Computer Communication (EIECC), pp. 1294–1298, Dec. 2024.